\documentclass[12pt]{article}
\usepackage[dvips]{epsfig}

\renewcommand{\thefootnote}{\fnsymbol{footnote}}

\begin{document}

 \thispagestyle{empty}
 \begin{flushright}
 {MZ-TH/01-32} \\[3mm]
 {hep-th/0311075} \\[3mm]
 {October 2003}
 \end{flushright}
 \vspace*{2.0cm}
 \begin{center}
 {\large \bf
   Explicitly symmetrical treatment of three-body phase space}
 \end{center}
 \vspace{1cm}
 \begin{center}
 A.~I.~Davydychev$^{a,b,c,}$\footnote{
Present address: Schlumberger, SPC, 155 Industrial Dr., Sugar Land,
TX~77478, USA.\\
Email address: davyd@thep.physik.uni-mainz.de} \ \ 
    and \ \ 
 R.~Delbourgo$^{a,}$\footnote{
Email address: bob.delbourgo@utas.edu.au}
\\
 \vspace{1cm}
$^{a}${\em School of Mathematics and Physics,
University of Tasmania,\\ GPO Box 252-21, Hobart,
Tasmania 7001, Australia}
\\
\vspace{.3cm}
$^{b}${\em Institute for Nuclear Physics, Moscow State University,\\
119992 Moscow, Russia}
\\
\vspace{.3cm}
$^{c}${\em Department of Physics, University of Mainz,\\
Staudingerweg 7, D-55099 Mainz, Germany}
\\
\end{center}
\hspace{3in}


\begin{abstract}
We derive expressions for three-body phase space that are 
{\em explicitly} symmetrical in the masses of the three particles.
We study geometrical properties of the variables involved
in elliptic integrals and demonstrate that it is convenient
to use the Jacobian zeta function to express the results
in four and six dimensions. 
\end{abstract}


\newpage
\renewcommand{\thefootnote}{\arabic{footnote}}
\setcounter{footnote}{0}

\section{Introduction} 
\setcounter{equation}{0}

The subject of three-body (and, in general, $N$-body)
relativistic phase space is as old as the hills
and one might well think that all that there is to know is known about it. 
In numerical and experimental terms this is indeed true: for a long time
Dalitz plots~\cite{Dalitz,Fabri} have been routinely used in picturing data 
and they prove extremely helpful for picking out resonant intermediate 
states of particular spin by their preferential population of the plots. In 
the absence of any amplitude modulation by resonances or otherwise, the 
plots are at their blandest as they just represent three-body phase space. 

One of the first comprehensive references on this subject
is the paper by Almgren~\cite{Almgren}. In his normalization,
the integral over the $N$-particle phase space is defined as
\begin{equation}
\label{def_I_N}
I_N^{(D)}(p, m_1, \ldots , m_N) = 
\int \cdots \int 
\left\{ \prod_{i=1}^N
{\rm d}^D p_i \; \delta(p_i^2-m_i^2)\; \theta(p_i^0) \right\}\;
\delta\left(\sum_{i=1}^N p_i - p \right) \; ,
\end{equation}
where $p$ is the total momentum. From now on, we will frequently
use the notation $p=\sqrt{p^2}$, since usually it is easy
to distinguish it from the cases when the four-dimensional vector $p$
is meant. As a rule, we will also omit the arguments of $I_N^{(D)}$.
In four dimensions ($D=4$), 
we will denote $I_N\equiv I_N^{(4)}$ (this is the original Almgren's
notation).
More details about integrals in other dimensions 
can be found in Ref.~\cite{BDR} and in the rest of this paper.
Worth mentioning is that $I_N^{(D)}$ is easy to work out for odd values of $D$,
whereas considering even values of $D$ brings in elliptic functions
and is more difficult. 

For kinematical reasons, it is clear
that the results for the integrals~(\ref{def_I_N}) have no
physical meaning if the absolute value 
of the momentum $p$ is less than the sum of the masses.
Therefore, in what follows we will imply that all results 
for $I_N^{(D)}$ are accompanied by 
$\theta\big\{p^2-(m_1+\ldots+m_N)^2\big\}$, without writing 
this theta function explicitly.
In Refs.~\cite{Hagedorn,Almgren}, integral recurrence relations for $I_N$
(at $D=4$) were discussed. For an arbitrary dimension $D$, the generating 
relation can be presented as
\begin{equation}
\label{Almgren_rec}
I_N^{(D)}(p,m_1,\ldots,m_N) = \int {\rm d}s \;
I_{R+1}^{(D)}(p,\sqrt{s},m_{N-R+1},\ldots,m_N)\;
I_{N-R}^{(D)}(\sqrt{s},m_1,\ldots,m_{N-R}) \; .
\end{equation}
Taking into account the theta functions associated with $I_{N-R}^{(D)}$ and
$I_{R+1}^{(D)}$, one can see that the actual limits of the integration
variable $s$
in Eq.~(\ref{Almgren_rec}) extend from $\left(\sum_{i=1}^{N-R}m_i\right)^2$
to $\left(p-\sum_{i=N-R+1}^{N}m_i\right)^2$.
Another type of integral recurrence relations for $I_N^{(D)}$, with respect 
to the value of $D$, was considered in Ref.~\cite{DR}.

The simplest example is the two-particle phase space, $N=2$.
In this case, the phase-space integral~(\ref{def_I_N}) 
in four dimensions can be easily evaluated as
\begin{equation}
\label{I_2}
I_2 = \frac{\pi}{2p^2}\sqrt{\lambda(p^2,m_1^2,m_2^2)} \; ,
\end{equation}
where 
\begin{equation}
\lambda(x,y,z)\equiv x^2+y^2+z^2-2xy-2yz-2zx
\label{Kaellen}
\end{equation} 
is nothing but the well-known K\"allen function~\cite{Kaellen}. 

Using Eqs.~(\ref{Almgren_rec}) and (\ref{I_2}) (for the case $D=4$, $R=1$), 
one can obtain the following integral representation~\cite{Almgren,BBBB}
for the three-particle ($N=3$) phase space:
\begin{eqnarray}
\label{I_3_sqrt}
I_3 =
\frac{\pi^2}{4p^2} \int\limits_{s_2}^{s_3} \frac{\mbox{d}s}{s}\;
\sqrt{(s-s_1)(s-s_2)(s_3-s)(s_4-s)} \; ,
\end{eqnarray}
with
\begin{equation}
\label{s_i}
s_1=(m_1-m_2)^2, \quad s_2=(m_1+m_2)^2, \quad s_3=(p-m_3)^2, \quad
s_4=(p+m_3)^2,
\end{equation}
so that $s_1\leq s_2\leq s_3 \leq s_4$. 
The result of the calculation of the integral~(\ref{I_3_sqrt})
can be expressed in terms of the elliptic integrals~\cite{Almgren,BBBB}
(for convenience, we collect the definitions and relevant
properties of elliptic integrals in an Appendix),
\begin{eqnarray}
\label{bbbb}
I_3
&=& \frac{\pi^2}{4 p^2\sqrt{Q_{+}}}
\Biggl\{ 
\frac{1}{2} Q_{+} (m_1^2 + m_2^2 + m_3^2 + p^2) E(k)
\nonumber \\ &&
+ 4 m_1 m_2 \left[ (p-m_3)^2 - (m_1-m_2)^2 \right]
\left[ (p+m_3)^2-m_3 p + m_1 m_2 \right] K(k)
\nonumber \\ &&
+ 8 m_1 m_2 
\left[ (m_1^2 + m_2^2)(p^2+m_3^2) - 2 m_1^2 m_2^2 - 2 m_3^2 p^2 \right]
\Pi\left( \alpha_1^2 , k \right)
\nonumber \\ &&
- 8 m_1 m_2 (p^2-m_3^2)^2
\Pi\left(\alpha_2^2, k \right)
\Biggl\} \; ,
\end{eqnarray}
where we use the following notations:
\begin{eqnarray}
Q_{+} &\!\!\equiv\!\!& 
(p\!+\!m_1\!+\!m_2\!+\!m_3)(p\!+\!m_1\!-\!m_2\!-\!m_3)
   (p\!-\!m_1\!+\!m_2\!-\!m_3)(p\!-\!m_1\!-\!m_2\!+\!m_3) \; ,
\nonumber \\
Q_{-} &\!\!\equiv\!\!& 
(p\!-\!m_1\!-\!m_2\!-\!m_3)(p\!-\!m_1\!+\!m_2\!+\!m_3)
   (p\!+\!m_1\!-\!m_2\!+\!m_3)(p\!+\!m_1\!+\!m_2\!-\!m_3) \; ,
\\
k &\equiv& \sqrt{\frac{Q_{-}}{Q_{+}}} \; , \quad
\alpha_1^2 = 
\frac{(p-m_3)^2-(m_1+m_2)^2}{(p-m_3)^2-(m_1-m_2)^2}\; ,
\quad
\alpha_2^2=\frac{(m_1-m_2)^2}{(m_1+m_2)^2}\alpha_1^2\; .
\end{eqnarray}
We note that in Ref.~\cite{BBBB} the notation
$q_{\pm \pm} \equiv (p\pm m_3)^2 - (m_1 \pm m_2)^2$ was used.
In particular, we have $Q_{+}=q_{++}q_{--}$,
$Q_{-}=q_{+-}q_{-+}$,
$k^2 = q_{+-} q_{-+}/(q_{++} q_{--})$,
$\alpha_1^2=q_{-+}/q_{--}$.
Note that $Q_{\pm}$ differ by the sign of $p$ only.

It is clear from the definition~(\ref{def_I_N}) that $I_3$ should 
be a symmetrical function of the three masses $m_1, m_2, m_3$.
The representation~(\ref{bbbb}) in terms of elliptic integrals is 
however {\em not} 
explicitly symmetrical in the masses, although it must be implicitly so. 
One may of course generate a symmetrical form by averaging the 
unsymmetrical-looking expressions over the three possible permutations
of $m_i$, but this would be ``cheating'' since each one of them should 
be symmetrical by itself, although this is hardly transparent.

Note that the quantities $Q_{+}$ and $Q_{-}$ (and, therefore,
the argument $k$)
are totally symmetric in $m_1$, $m_2$, $m_3$. (In fact, they are
symmetric in all four arguments $p, m_1, m_2, m_3$.)
Therefore, the term containing $E(k)$ in Eq.~(\ref{bbbb}) is also symmetric.
The function $K(k)$ itself is also symmetric, but its coefficient is not
symmetric. 
We also note that the product of $Q_{+}$ and $Q_{-}$ produces
the quantity
\begin{eqnarray}
D_{123} &\equiv& Q_{+} Q_{-} = 
\left[p^2-(m_1+m_2+m_3)^2\right]
\left[p^2-(-m_1+m_2+m_3)^2\right]
\nonumber \\ && \hspace*{17mm} \times
\left[p^2-(m_1-m_2+m_3)^2\right]
\left[p^2-(m_1+m_2-m_3)^2\right]
\end{eqnarray}
that occurs in recurrence relations for the sunset diagram
(see, e.g., in~\cite{Tarasov_sunset,DS}).
It should be noted that the imaginary part of the sunset diagram
is proportional to the three-particle phase-space integral.
For instance, in the notation of Ref.~\cite{BBBB}, 
$\mbox{Im}(T_{123})=-4\pi^{-1}\;I_3$. 
We also note that $\rho_N^D$ considered in Ref.~\cite{BDR,DR}
are related to $I_N^{(D)}$ as
$\rho_N^D = (2\pi)^{N+D-ND}\; I_N^{(D)}$.

For equal masses, $m_1=m_2=m_3\equiv m$, Eq.~(\ref{bbbb}) yields
\begin{equation}
\label{equal_m}
\frac{\pi^2}{4p^2}\; \sqrt{(p-m)(p+3m)}
\Biggl\{ \frac{1}{2} (p-m)(p^2+3m^2) E(k_{\rm eq}) 
- 4m^2 p K(k_{\rm eq}) \Biggl\} ,
\end{equation}
with
\begin{equation}
\label{k_eq}
k_{\rm eq} = \sqrt{\frac{(p+m)^3 (p-3m)}{(p-m)^3 (p+3m)}} \; .
\end{equation}
Some other special cases of Eq.~(\ref{bbbb}) are described 
in Refs.~\cite{BBBB,AT}. 

This paper is devoted to a new way of exhibiting the results in an 
{\em explicitly} symmetrical manner. To do this, we will employ
another integral representation for $I_3$, in terms of
Mandelstam variables $s, t, u$~\cite{Mandelstam} and the Kibble 
cubic $\Phi(s,t,u)$~\cite{Kibble}.
In particular, we will show that it is convenient to present the 
result~(\ref{bbbb}) in terms of Jacobi $Z$ function whose definition 
is given in the Appendix.

\section{Phase space integrals}
\setcounter{equation}{0}

As an illustration, let us demonstrate how the connection with 
the Dalitz picture can be derived directly from 
the definition~(\ref{def_I_N}).
The $D$-dimensional vector $p$ can be presented as $(p^0,{\mbox{\bf p}})$,
where ${\mbox{\bf p}}$ is the $(D-1)$-dimensional Euclidean vector
of space components. 
Without loss of generality, we can work in the center-of-mass frame,
$p=(p^0,{\mbox{\bf 0}})$.
Using the integral representation
\begin{equation}
\delta\left(\sum_{i=1}^N p_i - p \right)
= \frac{1}{(2\pi)^D}
\int {\rm d}^D x \; \exp\left\{ {\rm i}\sum_{i=1}^N (p_i x) 
- {\rm i}(p x) \right\} \; ,
\end{equation}
with $(p x)=p^0 x^0$, we get
\begin{equation}
I_N^{(D)} = \frac{1}{(2\pi)^D}
\int {\rm d}^D x \; e^{-{\rm i} p^0 x^0}
\left\{ \prod_{i=1}^N \int
{\rm d}^D p_i \; \delta(p_i^2-m_i^2)\; \theta(p_i^0) \;
e^{{\rm i} (p_i x)}\right\}\; .
\end{equation}
(Similar method was used in Ref.~\cite{ABKT}.)
Integrating over $(D-1)$-dimensional angles of $\mbox{\bf p}_i$ we get
\begin{equation} 
\int {\rm d}^D p_i \; \delta(p_i^2\!-m_i^2) \;
\theta(p_i^0) \; e^{{\rm i} (p_i x)}
= \frac{(2\pi)^{(D-1)/2}}{2\xi^{(D-3)/2}} \int\limits_0^{\infty}
\frac{\rho_i^{(D-1)/2} {\rm d}\rho_i}{\sqrt{\rho_i^2+m_i^2}} \; 
J_{(D-3)/2}(\rho_i \xi) \;
e^{{\rm i} x^0 \sqrt{\rho_i^2+m_i^2}} \; ,
\end{equation}
with $\rho_i\equiv |\mbox{\bf p}_i|$ and $\xi\equiv |\mbox{\bf x}|$.
In the four-dimensional case the Bessel function 
reduces to an elementary function,
$J_{1/2}(\rho_i\xi)=\left[2/(\pi\rho_i\xi)\right]^{1/2}\;\sin(\rho_i\xi)$.
We note an analogy with the calculation of Feynman integrals
in the coordinate space~\cite{Mendels,GKP}, when each massive propagator 
yields a (modified) Bessel function. 

Let us consider, for example, the two-particle phase space. Then,
the integration over $\xi$ gives us 
\[
\int\limits_0^{\infty} \xi {\rm d}\xi \; 
J_{\nu}(\rho_1\xi) \; J_{\nu}(\rho_2\xi)
= 2 \delta(\rho_1^2-\rho_2^2) \; , 
\]
with $\nu=(D-3)/2$, so that we can put $\rho_1=\rho_2\equiv\rho$,
whereas the integration over $x^0$ yields another delta function,
$\delta\left(p-\sqrt{\rho^2+m_1^2}-\sqrt{\rho^2+m_2^2}\right)$
in the center-of-mass frame.
The resulting integral
\begin{equation}
I_2^{(D)} = \frac{\pi^{(D-1)/2}}{2\Gamma\left(\frac{D-1}{2}\right)} 
\int\limits_0^{\infty} \frac{\rho^{D-2}\; {\rm d}\rho}
{\sqrt{\rho^2+m_1^2}\sqrt{\rho^2+m_2^2}}\;
\delta\left( p-\sqrt{\rho^2+m_1^2}-\sqrt{\rho^2+m_2^2}\right)
\end{equation}
can be easily evaluated, yielding (see, e.g., in Ref.~\cite{DR})
\begin{equation}
\label{I_2^D}
I_2^{(D)} = \frac{\pi^{(D-1)/2}}
{(2p)^{D-2}\;\Gamma\left(\frac{D-1}{2}\right)}
\left[ \lambda(p^2, m_1^2, m_2^2) \right]^{(D-3)/2} \; ,
\end{equation}
where $\lambda$ is defined in Eq.~(\ref{Kaellen}).
For $D=4$, Eq.~(\ref{I_2^D}) reduces to the well-known answer~(\ref{I_2}).

For the three-particle phase-space integral we get
\begin{equation}
I_3^{(D)} = 
\frac{2^{(D-7)/2}\pi^{D-2}}{\Gamma\left(\frac{D-1}{2}\right)}
\int\limits_0^{\infty} \frac{{\rm d}\xi}{\xi^{(D-5)/2}}
\int\limits_{-\infty}^{\infty} {\rm d}x^0 \; e^{-{\rm i} p^0 x^0}
\prod_{i=1}^3 \int\limits_0^{\infty} 
\frac{\rho_i^{(D-1)/2} {\rm d}\rho_i}{\sqrt{\rho_i^2+m_i^2}} \;
J_{(D-3)/2}(\rho_i \xi)\;
e^{{\rm i} x^0 \sqrt{\rho_i^2+m_i^2}} \; .
\end{equation}
Here we can integrate over $\xi$, using (see Ref.~\cite{PBMII})
\begin{equation}
\int\limits_0^{\infty} \frac{\rm d \xi}{\xi^{\nu-1}}\; 
J_{\nu}(\rho_1\xi)\; J_{\nu}(\rho_2\xi)\; J_{\nu}(\rho_3\xi)
= \frac{2\theta\{ -\lambda(\rho_1^2,\rho_2^2,\rho_3^2) \} 
[ -\lambda(\rho_1^2,\rho_2^2,\rho_3^2) ]^{\nu-1/2} }
{\pi^{1/2} \Gamma\left(\nu+{\textstyle{\frac{1}{2}}}\right) 
(8\rho_1\rho_2\rho_3)^{\nu}} \; 
\end{equation}
(with $\nu=(D-3)/2$), where $\lambda$ is
the K\"allen function (\ref{Kaellen}).
In fact, in our case, when all $\rho_i\geq0$,
\begin{equation}
\theta\{ -\lambda(\rho_1^2,\rho_2^2,\rho_3^2) \} =
\theta(\rho_1+\rho_2-\rho_3)\;
\theta(\rho_2+\rho_3-\rho_1)\;
\theta(\rho_3+\rho_1-\rho_2) \; ,
\label{Theta_rho}
\end{equation}
i.e., it equals 1 when one can compose a triangle with sides
$\rho_1$, $\rho_2$, $\rho_3$, and gives 0 otherwise
(cf.\ Eq.~(11) of Ref.~\cite{D-ep}).

Introducing notation $\sigma_i=\sqrt{\rho_i^2+m_i^2}$
and integrating over $x^0$ (getting a $\delta$ function)
we arrive at
\begin{eqnarray}
\label{I_3-3}
I_3^{(D)} &=& \frac{\pi^{D-2}}{\Gamma(D-2)} \int\limits_{m_1}^{\infty}
\int\limits_{m_2}^{\infty} \int\limits_{m_3}^{\infty}
{\rm d}\sigma_1 \; {\rm d}\sigma_2 \; {\rm d}\sigma_3 \;
\delta(p-\sigma_1-\sigma_2-\sigma_3)\;
\nonumber \\ &&
\times
\left[
-\lambda(\sigma_1^2\!-\!m_1^2, \sigma_2^2\!-\!m_2^2, \sigma_3^2\!-\!m_3^2)
\right]^{(D-4)/2}
\theta\left\{
-\lambda(\sigma_1^2\!-\!m_1^2, \sigma_2^2\!-\!m_2^2, \sigma_3^2\!-\!m_3^2) 
\right\} \; . \hspace*{7mm}
\end{eqnarray}
In four dimensions the factor $[-\lambda]^{(D-4)/2}$ disappears and,
geometrically,
we need to calculate a closed area 
on the plane $\sigma_1+\sigma_2+\sigma_3=p^0\equiv p$,
with the boundary of the figure described by
\begin{equation}
\lambda(\sigma_1^2-m_1^2, \sigma_2^2-m_2^2, \sigma_3^2-m_3^2)=0,
\quad \sigma_1+\sigma_2+\sigma_3=p \; .
\end{equation}

Furthermore, introducing Mandelstam-type variables
\begin{equation}
\label{s,t,u}
s = p^2 + m_3^2 - 2p\sigma_3, \quad
t = p^2 + m_1^2 - 2p\sigma_1, \quad
u = p^2 + m_2^2 - 2p\sigma_2 \; 
\end{equation}
satisfying
\begin{equation}
\label{s+t+u}
s+t+u = m_1^2 + m_2^2 + m_3^2 + p^2 \equiv w_0 \; ,
\end{equation}
one arrives at another integral representation (the limits
of integration are discussed below),
\begin{equation}
\label{I_3^D_Phi}
I_3^{(D)} = \frac{\pi^{D-2}}{4p^{D-2}\Gamma(D-2)}
\int\int\int {\rm d}s\; {\rm d}t\; {\rm d}u\; 
\delta(s+t+u-w_0) \;
\left[ \Phi(s,t,u) \right]^{(D-4)/2}
\theta\left\{ \Phi(s,t,u) \right\} \; ,
\end{equation}
where
\begin{equation}
\Phi(s,t,u) = - \frac{1}{16p^2}\; 
\lambda\left\{ \lambda(s,m_3^2,p^2), \lambda(t,m_1^2,p^2),
\lambda(u,m_2^2,p^2) \right\} \; 
\label{Phi_lambda}
\end{equation}
can also be written in a more familiar
Kibble cubic form~\cite{Kibble},
\begin{eqnarray}
\Phi(s,t,u) &=& stu - s (m_1^2 m_2^2 + p^2 m_3^2) 
                    - t (m_2^2 m_3^2 + p^2 m_1^2) 
                    - u(m_3^2m_1^2+p^2m_2^2) 
\nonumber \\ && 
+ 2 (m_1^2 m_2^2 m_3^2 + p^2 m_1^2 m_2^2 
     + p^2 m_2^2 m_3^2 + p^2 m_3^2 m_1^2) \; ,
\end{eqnarray}
provided that the condition~(\ref{s+t+u}) is satisfied.
In particular, in four dimensions we have
\begin{equation}
\label{I_3_Phi}
I_3 = \frac{\pi^2}{4p^2}
\int\int\int {\rm d}s\; {\rm d}t\; {\rm d}u\; 
\delta(s+t+u-w_0) \;
\theta\left\{ \Phi(s,t,u) \right\} \; .
\end{equation}

According to the definition~(\ref{s,t,u}) in terms of $\sigma_i$,
one can see that the maximal values of $s$, $t$ and $u$ 
(corresponding to the upper limits 
of integration in Eqs.~(\ref{I_3^D_Phi}) and~(\ref{I_3_Phi}))
are $s_{\rm max}=(p-m_3)^2$, $t_{\rm max}=(p-m_1)^2$ and 
$u_{\rm max}=(p-m_2)^2$. To define the minimal values of $s$, $t$ and $u$,
the familiar Dalitz--Kibble picture given in Fig.~1 is useful.
Due to the condition~(\ref{s+t+u}), the region of integration
is restricted by a triangle, with $s\geq~s_{\rm min}=(m_1+m_2)^2$, 
$t\geq~t_{\rm min}=(m_2+m_3)^2$ and $u\geq~u_{\rm min}=(m_1+m_3)^2$. 
Moreover, due to the theta function 
$\theta\left\{ \Phi(s,t,u) \right\}$
the region of integration is in fact restricted by the interior
of the cubic curve $\Phi(s,t,u)=0$, see Fig.~1.
Within that area,
the Mandelstam variables $s$, $t$ and $u$ take their minimal values
in points $O_s$, $O_t$ and $O_u$, respectively, whereas their 
maximal values correspond to the points $P_s$, $P_t$ and $P_u$.
(The dashed triangles will be discussed below in Section~4.)

\begin{figure}[t]
\refstepcounter{figure}
\addtocounter{figure}{-1}  
\begin{center}
\centerline{\vbox{\epsfysize=80mm \epsfbox{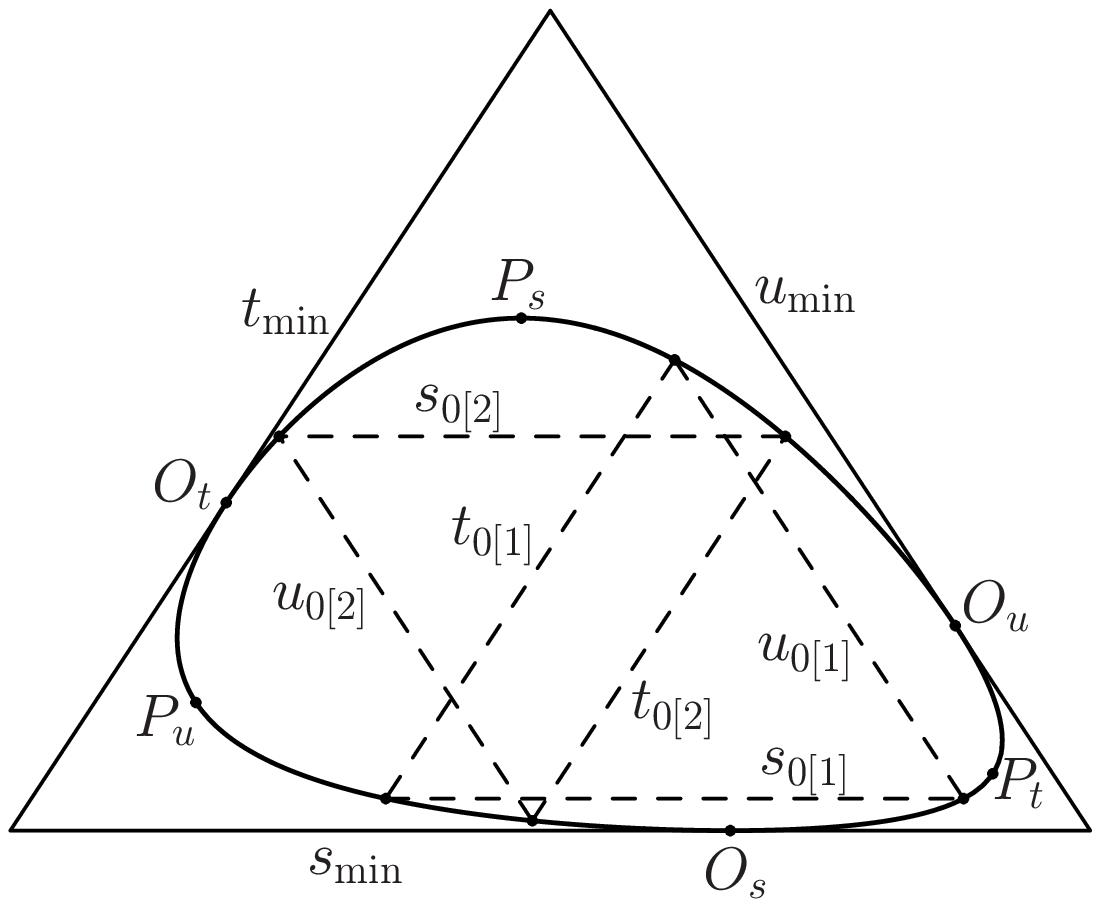}}}
\caption{The Dalitz--Kibble integration area}
\end{center}
\end{figure}

The function $\Phi(s,t,u)$ has a maximum within the region of
integration. For equal masses, the maximal value
$\Phi_{\rm max}=\frac{1}{27}\;p^2(p^2-9m^2)^2$
occurs at $s=t=u=\frac{1}{3}\;(p^2+3m^2)$. For the general unequal 
masses, one needs to solve a fourth-order algebraic equation to
find the position of the maximum.

We note that the representation~(\ref{Phi_lambda}) can be 
extracted from Eq.~(5.39) of~\cite{BK},
using symmetry properties. Our $\Phi(s,t,u)$
corresponds to $-G(s,t,p^2,m_2^2,m_1^2,m_3^2)$, in the
notations of~\cite{BK}. The $G$-function is symmetric
with respect to the permutations of three pairs of
arguments, $(s,t)$, $(p^2,m_2^2)$ and $(m_1^2,m_3^2)$.
Although the authors presume from their Eq.~(5.39) that
``from a practical point of view this identity is not
very useful'', we found that its symmetric form is certainly
helpful in understanding the structural properties of
phase space integrals.

\section{Geometrical interpretation}
\setcounter{equation}{0}

Let us introduce
\begin{equation}
c_{12} = \frac{s-m_1^2-m_2^2}{2 m_1 m_2}, \quad
c_{23} = \frac{t-m_2^2-m_3^2}{2 m_2 m_3}, \quad
c_{13} = \frac{u-m_1^2-m_3^2}{2 m_1 m_3} \; .
\end{equation}
Then, the function $\Phi(s,t,u)$ can be presented as a Gram determinant,
\begin{equation}
\label{Gram1}
\Phi(s,t,u) = 4 m_1^2 m_2^2 m_3^2 \;
\left|
\begin{array}{ccc}
1 & c_{12} & c_{13} \\
c_{12} & 1 & c_{23} \\
c_{13} & c_{23} & 1
\end{array}
\right| \; ,
\end{equation}
whereas the $\delta$ function becomes
\begin{eqnarray}
&& \delta(s+t+u-m_1^2-m_2^2-m_3^2-p^2)
\nonumber \\ &&
\hspace*{20mm}
\Rightarrow   
\delta\left(m_1^2+m_2^2+m_3^2 + 2 m_1 m_2 c_{12}
+ 2 m_2 m_3 c_{23} + 2 m_1 m_3 c_{13} -p^2 \right)\; .
\end{eqnarray}
In this way, we get 
\begin{eqnarray}
I_3 &=& \frac{2\pi^2}{p^2} m_1^2 m_2^2 m_3^2
\int\int\int {\rm d}c_{12}\; {\rm d}c_{13}\; {\rm d}c_{23}\;
\theta\left(
\left|
\begin{array}{ccc}
1 & c_{12} & c_{13} \\
c_{12} & 1 & c_{23} \\
c_{13} & c_{23} & 1
\end{array}
\right|
\right) \;
\nonumber \\ && \times
\delta\left(m_1^2+m_2^2+m_3^2 + 2 m_1 m_2 c_{12}
+ 2 m_2 m_3 c_{23} + 2 m_1 m_3 c_{13} -p^2 \right) \; ,
\end{eqnarray} 
where the integration extends over $c_{jl}\geq 1$.

\begin{figure}[t]
\refstepcounter{figure}
\addtocounter{figure}{-1}  
\begin{center}
\centerline{\vbox{\epsfysize=80mm \epsfbox{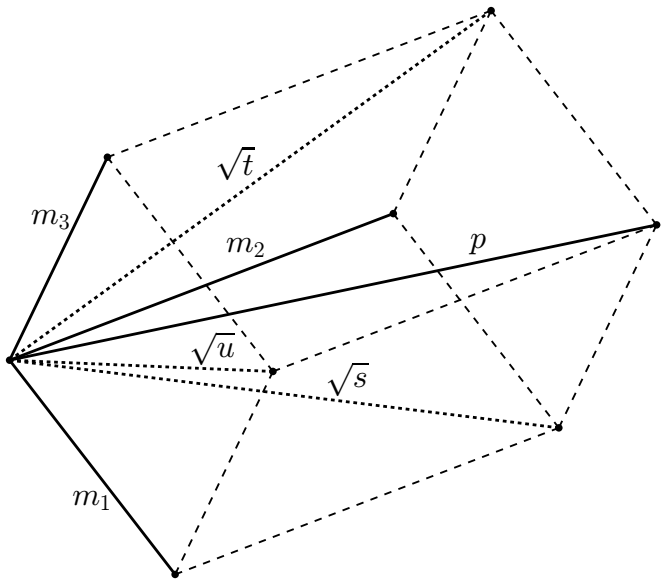}}}
\caption{The parallelepiped interpretation}
\end{center}
\end{figure}

If one were to interpret $c_{jl}$ as the cosines of the angles
between the $m_j$ and $m_l$ sides of a vertex of a parallelepiped
(formed by $m_1$, $m_2$ and $m_3$, see Fig.~2), then all these 
quantities would have a straightforward geometrical interpretation.
Namely, $\Phi(s,t,u)$ would be
$4\{ {\rm volume~of~parallelepiped} \}^2$,
whereas the $\delta$ function would tell us that the
``principal'' diagonal of this parallelepiped should
be equal to $p$. In this case, the quantities
$\sqrt{s}$, $\sqrt{t}$ and $\sqrt{u}$ could be identified as
the diagonals of the faces of the parallelepiped, see in Fig.~2.
Moreover, 
\begin{equation}
\label{sigmas}
\frac{p^2+m_1^2-t}{2pm_1}, \quad
\frac{p^2+m_2^2-u}{2pm_2} 
\quad {\rm and} \quad
\frac{p^2+m_3^2-s}{2pm_3}
\end{equation}
could be understood as cosines of
the angles between the diagonal $p$ and the $m_i$ sides
of the parallelepiped. In other words, these are
the angles between $p$ and $m_i$ in triangles
with sides
$\big(p,m_1,\sqrt{t}\big)$,
$\big(p,m_2,\sqrt{u}\big)$ and
$\big(p,m_3,\sqrt{s}\big)$, respectively.

Using this geometrical picture, we can mention a rather interesting
geometrical meaning of  Eq.~(\ref{Phi_lambda}). 
Namely, it tells us that the volume of the parallelepiped
is $(8/p)$ times the area of triangle whose sides are
given by the areas of triangles formed out of the principal
diagonal $p$, one of the face diagonals ($\sqrt{s}$, $\sqrt{t}$ or 
$\sqrt{u}$), and the appropriate $m_3$, $m_1$ or $m_2$ side.

However, when we are above the threshold, $p^2>(m_1+m_2+m_3)^2$,
the quantities $c_{jl}$ exceed one and therefore the expressions 
should be understood in the sense of analytic continuation,
i.e.\ as hyperbolic cosines.
The same is valid for the triangles $\left(p,m_3,\sqrt{s}\right)$,
etc.: they should also be understood in the
sense of analytic continuation, since $p\geq m_3+\sqrt{s}$,
etc. Therefore, the quantities
$\sigma_i/m_i$ 
should also be understood as hyperbolic cosines, 
whereas $\sqrt{\sigma_i^2-m_i^2}/m_i$ are hyperbolic sines.

Nevertheless, in the region
below the threshold (which we need, for instance, to describe the
real part of the sunset diagram), this geometrical picture
can have direct meaning, generalizing the picture we had
for the one-loop two-point function~\cite{DD}.

\section{Kibble cubic characteristics}
\setcounter{equation}{0}

Suppose 
\begin{equation}
\label{set}
(s_0, t_0, w_0-s_0-t_0), \quad 
(s_0, w_0-s_0-u_0, u_0), \quad
(w_0-t_0-u_0, t_0, u_0)
\end{equation}
all are the roots of the 
equation $\Phi(s,t,u)=0$. Then, we can present $\Phi(s,t,u)$ as
\begin{equation}
\label{Phi2}
\Phi(s,t,u) = s t u - s t_0 u_0 - s_0 t u_0 - s_0 t_0 u 
+ 2 s_0 t_0 u_0 \; .
\end{equation}
Furthermore, if we shift the Mandelstam variables as
\begin{equation}
s = s_0+s' \; , \qquad 
t = t_0+t' \; , \qquad
u = u_0+u' \; , 
\end{equation}
subject to the condition
\begin{equation}
s' + t' + u' = w_0 - s_0 - t_0 - u_0 \equiv w'_0 \; ,
\end{equation}
then
\begin{equation}
\label{Phi3}
\Phi(s,t,u) \Rightarrow 
s' t' u' + s_0 t' u' + s' t_0 u' + s' t' u_0 \; .
\end{equation}

Using equation Eq.~(\ref{Phi2}) and defining
\begin{equation}
c_{tu} \equiv \sqrt{\frac{t_0 u_0}{t u}} \; , \qquad
c_{st} \equiv \sqrt{\frac{s_0 t_0}{s t}} \; , \qquad
c_{su} \equiv \sqrt{\frac{s_0 u_0}{s u}} \; , 
\end{equation}
we arrive at another Gram determinant 
representation for $\Phi(s,t,u)$ (cf.\ Eq.~(\ref{Gram1})),
\begin{equation}
\label{determinant}
\Phi(s,t,u) = s t u \;
\left|
\begin{array}{ccc}
1 & c_{tu} & c_{st} \\
c_{tu} & 1 & c_{su} \\
c_{st} & c_{su} & 1 
\end{array}
\right| \; .
\end{equation} 

There are (at least) two sets of solutions~(\ref{set}) that
can be described as
\begin{equation}
\label{sets12}
s_0 = \frac{A_1 A_2}{A_3} \; , \qquad
t_0 = \frac{A_2 A_3}{A_1} \; , \qquad
u_0 = \frac{A_1 A_3}{A_2} \; , 
\end{equation}
so that Eq.~(\ref{Phi2}) yields
\begin{equation}
\Phi(s,t,u) = s t u - A_1^2 t - A_2^2 u - A_3^2 s 
+ 2 A_1 A_2 A_3 \; .
\end{equation}

{\em The first set} of solutions corresponds to
\begin{equation}
\label{set1a}
A_1 \equiv p m_1 + m_2 m_3 \; , \quad
A_2 \equiv p m_2 + m_3 m_1 \; , \quad
A_3 \equiv p m_3 + m_1 m_2 \; .
\end{equation}
For this set, we have
\begin{eqnarray}
&& w'_0 = w_0-s_0-t_0-u_0 =
\frac{m_1 m_2 m_3 p \; Q_{-}}{A_1 A_2 A_3} \; ,
\\ && 
c_{tu} = \frac{p m_3+m_1 m_2}{\sqrt{tu}}, \quad
c_{st} = \frac{p m_2+m_1 m_3}{\sqrt{st}}, \quad
c_{su} = \frac{p m_1+m_2 m_3}{\sqrt{su}} \; .
\end{eqnarray}
Note that if we change $p\to -p$ in Eq.~(\ref{set1a}), this would 
also be a solution, which would correspond to a ``non-physical'' 
branch of the Kibble cubic.

{\em The second set} of solutions corresponds to
\begin{equation}
\label{set2a}
A_1 \equiv {\textstyle{\frac{1}{2}}} (p^2+m_1^2-m_2^2-m_3^2) \; , \quad
A_2 \equiv {\textstyle{\frac{1}{2}}} (p^2-m_1^2+m_2^2-m_3^2) \; , \quad
A_3 \equiv {\textstyle{\frac{1}{2}}} (p^2-m_1^2-m_2^2+m_3^2) \; . 
\end{equation}
For this set, we get
\begin{eqnarray}
&& \hspace*{-10mm}
w'_0 = w_0-s_0-t_0-u_0 = 
- \frac{Q_{+} Q_{-}}{16 A_1 A_2 A_3} 
= - \frac{D_{123}}{16 A_1 A_2 A_3},
\\ && \hspace*{-10mm}
c_{tu} = \frac{p^2\!-\!m_1^2\!-\!m_2^2\!+\!m_3^2}{2\sqrt{tu}}, \quad\!\!
c_{st} = \frac{p^2\!-\!m_1^2\!+\!m_2^2\!-\!m_3^2}{2\sqrt{st}}, \quad\!\!
c_{su} = \frac{p^2\!+\!m_1^2\!-\!m_2^2\!-\!m_3^2}{2\sqrt{su}} .
\hspace*{7mm}
\label{cosines2}
\end{eqnarray}
It should be noted that the value of $s_0$ corresponding 
to the second set satisfies
\begin{equation}
\lambda(s_0, m_1^2, m_2^2) = \lambda(s_0, p^2, m_3^2) \; ,
\end{equation}
i.e., the areas (or their analytical continuations) 
of triangles with sides $(\sqrt{s_0},m_1,m_2)$
and $(\sqrt{s_0},p,m_3)$ are equal. Analogously,
\begin{equation}
\lambda(t_0, m_2^2, m_3^2) = \lambda(t_0, p^2, m_1^2) \; ,
\qquad
\lambda(u_0, m_1^2, m_3^2) = \lambda(u_0, p^2, m_2^2) \; .
\end{equation}
Moreover, one can get the direct geometrical interpretation
of the quantities (\ref{cosines2}) through the familiar 
parallelepiped picture given in Fig.~2. Namely, $c_{tu}$ is nothing 
but the cosine between the face diagonals $\sqrt{t}$ and
$\sqrt{u}$. Accordingly, $c_{su}$ is the cosine 
of the angle between $\sqrt{s}$ and $\sqrt{u}$ diagonals,
whilst $c_{st}$ is the cosine of the angle between
$\sqrt{s}$ and $\sqrt{t}$ diagonals.
If we construct a tetrahedron using the $\sqrt{s}$, $\sqrt{t}$
and $\sqrt{u}$ diagonals then, according to 
Eq.~(\ref{determinant}),
$\Phi(s,t,u)$ would represent 36 times its volume squared. 

In the Dalitz--Kibble plot shown in Fig.~1 we connect 
the points~(\ref{set}) 
for each of the two sets by dashed lines, 
introducing subscripts {\footnotesize{$[1]$}} and 
{\footnotesize{$[2]$}}
for the first and the second set, respectively. 
The two resulting ``dashed'' triangles 
indicate that the two sets of solutions are complementary
to each other. 
Namely, the boundary of the Dalitz plot 
confines the products $tu$, $st$ and $su$ as follows:
\begin{eqnarray}
\label{confine}
(p m_3+ m_1 m_2)^2 \leq tu \leq 
  {\textstyle{\frac{1}{4}}} (p^2-m_1^2-m_2^2+m_3^2)^2 \; ,
\nonumber \\
(p m_2+ m_1 m_3)^2 \leq st \leq 
  {\textstyle{\frac{1}{4}}} (p^2-m_1^2+m_2^2-m_3^2)^2 \; ,
\nonumber \\
(p m_1+ m_2 m_3)^2 \leq su \leq 
  {\textstyle{\frac{1}{4}}} (p^2+m_1^2-m_2^2-m_3^2)^2 \; ,
\end{eqnarray}
or, equivalently,
\begin{equation}
\label{confine2}
(t_0 u_0)_{[1]} \leq tu \leq 
   (t_0 u_0)_{[2]}  \; , \quad
(s_0 t_0)_{[1]} \leq st \leq 
   (s_0 t_0)_{[2]}  \; , \quad
(s_0 u_0)_{[1]} \leq su \leq 
   (s_0 u_0)_{[2]}  \; .
\end{equation}
In other words, the first and the second sets yield, respectively, 
the minimal and the maximal values of $tu$, $st$ and $su$.

Let us consider the corresponding values of the ``cosines''
$c_{su}$, $c_{st}$ and $c_{tu}$. For the first set, 
$c_{su}$, $c_{st}$ and $c_{tu}$
would vary between 1 and $\cos\varphi_i$ ($i=1,2,3$), 
respectively, where 
\begin{equation}
\label{cos_phi}
\cos\varphi_1 = \frac{2(p m_1-m_2 m_3)}{p^2\!+\!m_1^2\!-\!m_2^2\!-\!m_3^2} , 
\quad\!
\cos\varphi_2 = \frac{2(p m_2-m_3 m_1)}{p^2\!-\!m_1^2\!+\!m_2^2\!-\!m_3^2} , 
\quad\!
\cos\varphi_3 = \frac{2(p m_3-m_1 m_2)}{p^2\!-\!m_1^2\!-\!m_2^2\!+\!m_3^2} .
\end{equation}
For the second set, $c_{su}$, $c_{st}$ and $c_{tu}$
would vary between 1 and $1/\cos\varphi_i$. This means
that we need to understand them in the sense of 
analytic continuation.  

The angles $\varphi_i$ will be very important below. 
Their sines can be presented as
\begin{equation}
\label{sin_phi}
\sin\varphi_1 =  
\frac{\sqrt{Q_{+}}}{p^2\!+\!m_1^2\!-\!m_2^2\!-\!m_3^2} , 
\quad\!
\sin\varphi_2 =  
\frac{\sqrt{Q_{+}}}{p^2\!-\!m_1^2\!+\!m_2^2\!-\!m_3^2} ,
\quad\!
\sin\varphi_3 =  
\frac{\sqrt{Q_{+}}}{p^2\!-\!m_1^2\!-\!m_2^2\!+\!m_3^2} .
\end{equation}
It is interesting that the corresponding Gram determinant 
can be factorized as
\begin{eqnarray}
\label{Gram_phi_i}
\left| 
\begin{array}{ccc}
1 & -\cos\varphi_3 & -\cos\varphi_2 \\
-\cos\varphi_3 & 1  & -\cos\varphi_1 \\
-\cos\varphi_2 & -\cos\varphi_1 & 1 
\end{array}                        
\right|
= \frac{1}{k^2}\; \sin^2\varphi_1 \; \sin^2\varphi_2 \;
\sin^2\varphi_3 \; .
\end{eqnarray}
Eq.~(\ref{Gram_phi_i}) can be used to express $k$ 
in terms of $\varphi_i$. We also note that
\begin{equation}
\tan\frac{\varphi_3}{2} 
= \sqrt{\frac{(p-m_3)^2-(m_1-m_2)^2}{(p+m_3)^2-(m_1+m_2)^2}} \; ,
\end{equation}
and similarly for $\varphi_1$ and $\varphi_2$.
In particular,
one can see that at the threshold, $p=m_1+m_2+m_3$,
the angles $\varphi_i$ are related to the angles $\theta_i$
from Eq.~(20) of Ref.~\cite{BDU} (see also in Ref.~\cite{DS}) 
as $\varphi_i=\pi-2\theta_i$, and
\begin{equation}
\left. \left( \varphi_1 + \varphi_2 
+ \varphi_3 \right)\right|_{p=m_1+m_2+m_3} = \pi \; .
\end{equation}

We can also consider associated angles $\psi_i$, such that
\begin{equation}
\sin\psi_i = k \sin\varphi_i, \qquad
\cos\psi_i = \sqrt{1-k^2\sin^2\varphi_i} \; .
\end{equation}
Explicitly, we get
\begin{equation}
\sin\psi_3 = \frac{\sqrt{Q_{-}}}{p^2-m_1^2-m_2^2+m_3^2} ,
\qquad
\cos\psi_3 = \frac{2(p m_3+m_1 m_2)}{p^2-m_1^2-m_2^2+m_3^2} ,
\end{equation}
etc. For these angles, we get
\begin{eqnarray}
\left|
\begin{array}{ccc}
1 & \cos\psi_3 & \cos\psi_2 \\
\cos\psi_3 & 1  & \cos\psi_1 \\
\cos\psi_2 & \cos\psi_1 & 1
\end{array}
\right|
= \frac{1}{k^2}\; \sin^2\psi_1 \; \sin^2\psi_2 \;
\sin^2\psi_3 \; .
\end{eqnarray}

\section{A naturally symmetric representation}
\setcounter{equation}{0}

Using the representation~(\ref{Phi2})
for $\Phi(s,t,u)$, in terms of $s_0$, $t_0$ and $u_0$,
the three-body phase-space integral can be written as
\begin{equation}
\label{delta*theta}
I_3=\frac{\pi^2}{4p^2}
\int\!\!\int\!\!\int {\rm d}s\; {\rm d}t\; {\rm d}u\;
\delta(s+t+u-w_0)\;
\theta(s t u - s t_0 u_0 - s_0 t u_0 - s_0 t_0 u + 2 s_0 t_0 u_0) \;, 
\end{equation}
with $w_0=p^2+m_1^2+m_2^2+m_3^2$.
Integrating over $u$ yields
\begin{equation}
I_3=\frac{\pi^2}{4p^2}
\int\!\!\int {\rm d}s\; {\rm d}t\;
\theta\bigl\{ (st - s_0 t_0) (w_0-s-t) - s t_0 u_0 - s_0 t u_0 
+ 2 s_0 t_0 u_0 \bigl\} \; .
\end{equation}
Then, integrating over $t$, we basically obtain the difference
between the roots of the quadratic argument of the $\theta$ function,
which is
\[
\frac{1}{s}
\sqrt{s^4\!-\!2w_0s^3 \!+\! (w_0^2\!+\!2 s_0t_0\!+\!2s_0u_0\!-\!4t_0u_0)s^2
\!-\!2(w_0t_0\!+\!w_0u_0\!-\!4u_0t_0)s_0 s \!+\! s_0^2(t_0\!-\!u_0)^2} .
\]
It is easy to check that for both sets of $(s_0,t_0,u_0)$ the square
root takes the familiar form~(\ref{I_3_sqrt}),
which yields the non-symmetric result~(\ref{bbbb}) in terms 
of elliptic integrals.

Starting from the representation (\ref{I_3^D_Phi}), one can easily generalize 
the result~(\ref{I_3_sqrt}) to the $D$-dimensional case as
\begin{equation}
\label{I_3^D_sqrt}
I_3^{(D)} = \frac{\pi^{D-1}}
                 {(4p)^{D-2} \Gamma^2\left(\frac{D-1}{2}\right)}
\int\limits_{s_2}^{s_3} \frac{{\rm d}s}{s^{D/2-1}}\;
\left[(s\!-\!s_1)(s\!-\!s_2)(s_3\!-\!s)(s_4\!-\!s)\right]^{(D-3)/2} \; ,
\end{equation}
with $s_i$ given in Eq.~(\ref{s_i}). Another way to derive 
the representation~(\ref{I_3^D_sqrt}) is to use the recurrence
relation~(\ref{Almgren_rec}), 
\begin{equation}
\label{Almgren_rec3}
I_3^{(D)} = \int\limits_{s_2}^{s_3} {\rm d}s\;
I_2^{(D)}(p,\sqrt{s},m_3)\; I_2^{(D)}(\sqrt{s},m_1,m_2)\; ,
\end{equation}
and substitite the result~(\ref{I_2^D}) for $I_2^{(D)}$.
The result~(\ref{I_3^D_sqrt}) corresponds to Eq.~(9) of Ref.~\cite{BDR}.
[We note that the overall factor on the r.h.s.\ 
of Eq.~(9) of Ref.~\cite{BDR} should be corrected: 
$(32\pi)^{2-2\ell}$ should be changed into $\frac{1}{2}(16\pi)^{2-2\ell}$,
with $\ell=D/2$].

Using representation~(\ref{I_3^D_sqrt}), it is
easy to see (just substituting $s=x^2$) that all {\em odd}-dimensional 
phase-space integrals can be expressed in terms of 
polynomial functions (see, e.g., in Refs.~\cite{Rajantie,GKP,Roberts}),
\begin{eqnarray}
I_3^{(3)} &=& \frac{\pi^2}{2p} (p-m_1-m_2-m_3) \; ,
\\
I_3^{(5)} &=& \frac{\pi^4}{60p^3} (p-m_1-m_2-m_3)^3
\Biggl[
\frac{1}{7} (p-m_1-m_2-m_3)^4 + (m_1+m_2+m_3) p^3 
\nonumber \\ &&
     - 2 (m_1^2+m_2^2+m_3^2) p^2 + (m_1^3+m_2^3+m_3^3) p
     + 12 m_1 m_2 m_3 p
\nonumber \\ &&  
     - (m_1\!+\!m_2\!+\!m_3) (m_1\!+\!m_2) (m_2\!+\!m_3) (m_3\!+\!m_1)
     + 4 m_1 m_2 m_3 (m_1\!+\!m_2\!+\!m_3) \Biggl] \; ,
\hspace*{8mm}
\end{eqnarray}
etc., which are explicitly symmetric in the masses $m_i$. 
However, the results in {\em even} dimensions appear to be less trivial.

It is instructive to consider the two-dimensional case, $D=2$.
Then, the integral~(\ref{I_3^D_sqrt}) yields just the elliptic 
integral $K(k)$,
\begin{equation}
\label{I_3^2}
I_3^{(2)} = 
\int\limits_{s_2}^{s_3} \frac{{\rm d}s}
{\sqrt{(s\!-\!s_1)(s\!-\!s_2)(s_3\!-\!s)(s_4\!-\!s)}}
= \frac{2}{\sqrt{Q_{+}}}\; K(k) \; .
\end{equation}
This is of course explicitly symmetric in the masses without further ado. 
On the other hand, using the $\delta$ function in Eq.~(\ref{delta*theta}),
we can insert $1=(s+t+u)/w_0$ in the integrand, and then consider
the three resulting terms (with $s$, $t$ and $u$) separately.
In this way, we arrive at an alternative expression,
\begin{eqnarray}
\label{I_3^2_}
I_3^{(2)} &=& \frac{1}{w_0}\Biggl\{
\int\limits_{s_2}^{s_3} \frac{s\;{\rm d}s}
{\sqrt{(s\!-\!s_1)(s\!-\!s_2)(s_3\!-\!s)(s_4\!-\!s)}}
+ \int\limits_{t_2}^{t_3} \frac{t\;{\rm d}t}
{\sqrt{(t\!-\!t_1)(t\!-\!t_2)(t_3\!-\!t)(t_4\!-\!t)}}
\nonumber \\ &&
+\int\limits_{u_2}^{u_3} \frac{u\;{\rm d}u}
{\sqrt{(u\!-\!u_1)(u\!-\!u_2)(u_3\!-\!u)(u_4\!-\!u)}}
\Biggl\} \; ,
\end{eqnarray}
where the roots $t_i$ and $u_i$ can be obtained from $s_i$ given 
in Eq.~(\ref{s_i}) by proper permutation of the masses $m_i$.
Each of the integrals involved in Eq.~(\ref{I_3^2_}) can be expressed
in terms of Jacobian $Z$ function (see in Appendix). For example,
\begin{eqnarray}
\int\limits_{s_2}^{s_3} \frac{s\;{\rm d}s}
{\sqrt{(s\!-\!s_1)(s\!-\!s_2)(s_3\!-\!s)(s_4\!-\!s)}}
= \frac{\sin\varphi_3}{\sin\varphi_1\; \sin\varphi_2}\; K(k)
-K(k)\; Z(\varphi_3, k) \; ,
\end{eqnarray}
where $\varphi_i$ are nothing but the three angles 
defined in Eqs.~(\ref{cos_phi})--(\ref{sin_phi}). 
Comparing the resulting expression with the original result~(\ref{I_3^2})
we obtain a very useful relation between the three $Z(\varphi_i,k)$ 
functions,
\begin{equation}
Z(\varphi_1,k)+Z(\varphi_2,k)+Z(\varphi_3,k)
= k^2 \; \sin{\varphi_1} \; \sin{\varphi_2} \; \sin{\varphi_3} \; .
\label{Z+Z+Z}
\end{equation}

Let us now consider the four-dimensional integral $I_3^{(4)}\equiv I_3$,
namely, its representation~(\ref{I_3_sqrt}).
A useful observation is that the result would be simpler if we managed 
to get rid of $s$ in the denominator. In particular, it would contain
just one elliptic integral $\Pi$, rather than two.
How are we to eliminate $s$ in the denominator? Again,
using the $\delta$ function in Eq.~(\ref{delta*theta}),
we can insert $1=(s+t+u)/w_0$ in the integrand, and then consider
the three resulting terms (with $s$, $t$ and $u$) separately.
For the term with $s$, we perform the $t$ and $u$ integrations
and arrive at the same integral as in Eq.~(\ref{I_3_sqrt}), but without $s$
in the denominator. In two other integrals, we just integrate
in a different order, leaving as the last one the $t$ or $u$ integration, 
respectively. In this way, we obtain for the integral~(\ref{delta*theta})
\begin{eqnarray}
\label{delta*theta2}
&& \hspace*{-6mm}
\frac{\pi^2}{4p^2 w_0}
\int\!\!\int\!\!\int {\rm d}s\; {\rm d}t\; {\rm d}u\;
(s\!+\!t\!+\!u)\; \delta(s\!+\!t\!+\!u\!-\!w_0)\;
\theta(s t u - s t_0 u_0 - s_0 t u_0 - s_0 t_0 u + 2 s_0 t_0 u_0)  
\nonumber \\ 
&=& 
\frac{\pi^2}{4p^2 w_0} \Biggl\{ 
\int\limits_{s_2}^{s_3} \mbox{d}s\;
\sqrt{(s-s_1)(s-s_2)(s_3-s)(s_4-s)} 
+ \int\limits_{t_2}^{t_3} \mbox{d}t\;
\sqrt{(t-t_1)(t-t_2)(t_3-t)(t_4-t)}
\nonumber \\ && \hspace*{15mm}
+ \int\limits_{u_2}^{u_3} \mbox{d}u\;
\sqrt{(u-u_1)(u-u_2)(u_3-u)(u_4-u)}
\Biggl\} \; ,
\end{eqnarray} 
where, as before, the roots $t_i$ and $u_i$ can be obtained from $s_i$ given 
in Eq.~(\ref{s_i}) by permutation of the masses.

Using the formulae given in Ref.~\cite{Byrd} along with 
Eqs.~(\ref{caseIII}) and (\ref{141.01}), 
the $s$-integral in Eq.~(\ref{delta*theta2})
can be calculated in terms of a Jacobian $Z$ function (see Appendix),
\begin{eqnarray}
\label{sqrt_Z3}
&& \hspace*{-15mm}
\int\limits_{s_2}^{s_3} \mbox{d}s\;
\sqrt{(s-s_1)(s-s_2)(s_3-s)(s_4-s)}
\nonumber \\ 
&=& \sqrt{Q_{+}}
\Biggl\{ 2 (p^2 m_3^2 - m_1^2 m_2^2) K(k) \frac{Z(\varphi_3,k)}{\sin{\varphi_3}}
+ \frac{1}{6} Q_{-} K(k)
\nonumber \\ &&
+\frac{1}{6}
\left[
(p^2-m_1^2-m_2^2+m_3^2)^2 + 8 (p^2 m_3^2 + m_1^2 m_2^2)
\right]
\left[ E(k) - K(k) \right]
\Biggl\} \; ,
\end{eqnarray}
where $\varphi_3$ is one of the three angles 
defined in Eqs.~(\ref{cos_phi})--(\ref{sin_phi}). 

Collecting the results for all three integrals and using 
the relation~(\ref{Z+Z+Z}), we arrive at the symmetric result 
\begin{eqnarray}
\label{symmetric}
I_3 &=&  
\frac{\pi^2}{8 p^2} \Biggl\{ \sqrt{Q_{+}}
(p^2+m_1^2+m_2^2+m_3^2) \left[ E(k) - K(k) \right]
\nonumber \\ &&
+ Q_{+} K(k)
\left[
\frac{Z(\varphi_1,k)}{\sin^2{\varphi_1}}
+\frac{Z(\varphi_2,k)}{\sin^2{\varphi_2}}
+\frac{Z(\varphi_3,k)}{\sin^2{\varphi_3}}
\right]
\Biggl\} \; .
\end{eqnarray} 
This symmetric result can also be presented in terms of the elliptic 
integrals $\Pi$, using (see in \cite{Byrd})
\begin{equation}
\label{Z_Pi}
K(k)\; Z(\varphi_i,k) = \cot\varphi_i\; \sqrt{1-k^2\sin^2\varphi_i}
\left[ \Pi(k^2\sin^2\varphi_i, k) - K(k) \right] \; .
\end{equation}

In principle, one can also derive the result~(\ref{symmetric})
directly from the non-symmetric representation~(\ref{bbbb}) 
(see in Ref.~\cite{DD-ARC}),
in a tedious way relying on the use of several relations collected 
in the Appendix, including the addition formula~(\ref{Z_addition})
for Jacobi $Z$ functions. 

Worth noting is that in a similar
way one can obtain results for higher even dimensions $D$.
For instance, in six dimensions we get
\begin{eqnarray}
\label{I_3^6}
I_3^{(6)}
&=& \frac{\pi^4}{144p^4}\Biggl\{
\frac{Q_{+}^{1/2}}{20} \left[ E(k) \!-\! K(k) \right]
\Bigl[ 192 (p^8\!+\!m_1^8\!+\!m_2^8\!+\!m_3^8)
         - 112 (p^4\!+\!m_1^4\!+\!m_2^4\!+\!m_3^4)^2 
\nonumber \\ &&
        - 6(p^2\!+\!m_1^2\!+\!m_2^2\!+\!m_3^2)^4
        - 156 (p^6\!+\!m_1^6\!+\!m_2^6\!+\!m_3^6) 
             (p^2\!+\!m_1^2\!+\!m_2^2\!+\!m_3^2) \hspace*{-7mm}
\nonumber \\ && 
        +83 (p^4\!+\!m_1^4\!+\!m_2^4\!+\!m_3^4) 
            (p^2\!+\!m_1^2\!+\!m_2^2\!+\!m_3^2)^2  
\Bigl]
\nonumber \\ && 
+ \frac{1}{40} Q_{-} Q_{+}^{1/2} K(k)
\Bigl[ 3 (p^2+m_1^2+m_2^2+m_3^2)^2-16 (p^4+m_1^4+m_2^4+m_3^4) \Bigl]
\nonumber \\ && 
+ \frac{3}{4}\; 
\frac{Q_{+}^{5/2}\; K(k) }{\sin\varphi_1\; \sin\varphi_2\; \sin\varphi_3} 
\left[ \frac{Z(\varphi_1,k)}{\sin^2\varphi_1}
+ \frac{Z(\varphi_2,k)}{\sin^2\varphi_2}
+ \frac{Z(\varphi_3,k)}{\sin^2\varphi_3} \right]
\nonumber \\ && 
-\frac{3}{8} Q_{+}^2 (p^2+m_1^2+m_2^2+m_3^2)
K(k) \left[ \frac{Z(\varphi_1,k)}{\sin^4\varphi_1}
+ \frac{Z(\varphi_2,k)}{\sin^4\varphi_2}
+ \frac{Z(\varphi_3,k)}{\sin^4\varphi_3} \right] 
\Biggl\} \; .
\end{eqnarray}
As an alternative way to obtain results for higher values of $D$,
the approach of the paper~\cite{Tarasov_sunset} may be used. 

In the equal-mass case,
\begin{equation}
\varphi_1 = \varphi_2 = \varphi_3 \equiv \varphi_{\rm eq} \; , \quad
\sin\varphi_{\rm eq} = \frac{\sqrt{(p-m)(p+3m)}}{p+m}\; , \quad
\cos\varphi_{\rm eq} = \frac{2m}{p+m} \; .
\end{equation}
Here, using Eq.~(\ref{Z+Z+Z}) we get
\begin{equation}
\label{Z_eq}
Z(\varphi_{\rm eq}, k_{\rm eq}) 
= \frac{1}{3}\; k_{\rm eq}^2\; \sin^3\varphi_{\rm eq} \; ,
\end{equation}
with $k_{\rm eq}$ defined in Eq.~(\ref{k_eq}). 
In this way, we reproduce Eq.~(\ref{equal_m}), whereas for $D=6$ 
Eq.~(\ref{I_3^6}) yields
\begin{eqnarray}
\label{I_3^6_eq}
\frac{\pi^4}{2880p^4} \sqrt{(p-m)^3 (p+3m)}
\Bigl\{ && \!\! (p^4-9m^4) (p^4-42 p^2 m^2 + 9 m^4) 
\left[ E(k_{\rm eq})- K(k_{\rm eq})\right]
\nonumber \\ &&
+(p+m)^3 (p-3m) (p^4-36 p^2 m^2 + 27 m^4) K(k_{\rm eq}) 
\Bigl\} \; . \hspace*{8mm}
\end{eqnarray}
We note that Eq.~(\ref{Z_eq}) yields a reduction
formula of $Z(\varphi,k)$, for a special case when 
\[
k = \frac{\sqrt{1-2\cos\varphi}}{\sin\varphi (1-\cos\varphi)} \; .
\]

Another interesting limit corresponds to the case when one of
the masses vanishes (for example, $m_3\to0$). This corresponds to the case
$k\to1$, when $E(k)$ is finite ($E(1)=1$) whereas $K(k)$ develops 
logarithmic singularity.
At $m_3=0$, $\cos\varphi_3<0$ and $\varphi_3>\pi/2$, so that
we need to use Eq.~(\ref{Zeta_continued}). Using equations listed
in Ref.~\cite{Byrd}, we get
\begin{equation}
\label{k_to_1}
\lim\limits_{k\to1}\left\{ K(k) \left[ \pm Z(\varphi,k) 
- \sin\varphi\right] \right\}
= -\frac{1}{2} 
\ln\left( \frac{1+\sin\varphi}{1-\sin\varphi}\right) \; ,
\end{equation}
where plus or minus should be used for $\varphi<\pi/2$ or $\varphi>\pi/2$,
respectively. Let us consider Eq.~(\ref{symmetric}).
Using Eqs.~(\ref{k_to_1}) and (\ref{sin_phi}) we see that 
singular terms containing $K(k)$ cancel, and we arrive at the following result:
\begin{eqnarray}
\lim\limits_{m_3\to0} I_3 = \frac{\pi^2}{8p^2}
\Biggl\{ && \!\!\!\! \sqrt{Q_{+}} (p^2+m_1^2+m_2^2)
+\frac{1}{2}(p^2-m_1^2-m_2^2)^2\; 
\ln\left(\frac{p^2-m_1^2-m_2^2+\sqrt{Q_{+}}}
              {p^2-m_1^2-m_2^2-\sqrt{Q_{+}}} \right)
\nonumber \\ &&
- \frac{1}{2}(p^2+m_1^2-m_2^2)^2\; 
\ln\left(\frac{p^2+m_1^2-m_2^2+\sqrt{Q_{+}}}
              {p^2+m_1^2-m_2^2-\sqrt{Q_{+}}} \right)
\nonumber \\ &&
- \frac{1}{2}(p^2-m_1^2+m_2^2)^2\; 
\ln\left(\frac{p^2-m_1^2+m_2^2+\sqrt{Q_{+}}}
              {p^2-m_1^2+m_2^2-\sqrt{Q_{+}}} \right)
\Biggl\} \; ,
\end{eqnarray}
where $Q_{+}=\lambda(p^2,m_1^2,m_2^2)$ in this limit. 
It is easy to check that this expression is equivalent to known results
(see, e.g., in Refs.~\cite{Almgren,BBBB}). The advantage of our
approach is that the symmetry with respect to any of the remaining masses 
is always explicit, whereas non-symmetric expressions like Eq.~(\ref{bbbb})
lead to the answers which are not explicitly symmetric 
(cf.\ Eq.~(57) of Ref.~\cite{BBBB}).

\section{Conclusion}
\setcounter{equation}{0}

We have considered several representations for the three-particle
phase space, exploring their symmetry properties and 
geometrical meaning. It was shown that the angles $\varphi_i$
defined in Eqs.~(\ref{cos_phi})--(\ref{sin_phi}) are convenient
to describe the results for the three-particle phase-space integral~$I_3$.
In terms of the Jacobian $Z$ function (related 
to the elliptic integral $\Pi$ through Eq.(\ref{Z_Pi})), the result for $I_3$ 
in four dimensions is given in Eq.~(\ref{symmetric}). 
It is very compact and explicitly symmetric with respect 
to all masses $m_i$. Note that the three zeta functions $Z(\varphi_i,k)$
are connected through the relation~(\ref{Z+Z+Z}). This relation
can be obtained by comparing the representation~(\ref{I_3^2}) 
for two-dimensional integral $I_3^{(2)}$ with another representation
obtained by using the delta function properties.
We have also considered the six-dimensional case.
The result for $I_3^{(6)}$ is given in Eq.~(\ref{I_3^6}), also expressed
in terms of $Z(\varphi_i,k)$. 

\vspace{4mm}

{\bf Acknowledgements.}
We are pleased to acknowledge financial support from the Australian Research
Council under grant number~A00000780. 
Partical support from the Deutsche Forschungsgemeinschaft (A.~D.)
is also acknowledged.

\appendix
\section*{Appendix: Elliptic integrals}
\setcounter{equation}{0}
\renewcommand{\thesection}{A}

The normal elliptic integrals of the first and second kind 
are defined as
\begin{eqnarray}  
\label{elliptic_F_}
F(\varphi,k) &=& \int\limits_0^{\sin\varphi}
\frac{\mbox{d}t}{\sqrt{(1-t^2)(1-k^2t^2)}}
= \int\limits_0^{\varphi} \frac{\mbox{d}\psi}{\sqrt{1-k^2\sin^2\psi}}
\\
\label{elliptic_E_}
E(\varphi,k) &=& \int\limits_0^{\sin\varphi} 
\mbox{d}t \; \sqrt{\frac{1-k^2t^2}{1-t^2}}
= \int\limits_0^{\varphi} \mbox{d}\psi\; \sqrt{1-k^2\sin^2\psi} \; .
\end{eqnarray}
At $\varphi=\pi/2$ we get the complete elliptic integrals,
\begin{eqnarray}
\label{elliptic_K}
K(k) &=& F\left({\textstyle{\frac{\pi}{2}}},k\right) =
\int\limits_0^1 \frac{\mbox{d}t}{\sqrt{(1-t^2)(1-k^2t^2)}}
      = \frac{\pi}{2}\; \left. _2F_1\left(
      \begin{array}{c} -\frac{1}{2}, \; \frac{1}{2} \\ 1 \end{array}
      \right| k^2 \right) \; ,
\\
\label{elliptic_E}
E(k) &=& E\left({\textstyle{\frac{\pi}{2}}},k\right) =
\int\limits_0^1 \mbox{d}t \; \sqrt{\frac{1-k^2t^2}{1-t^2}}
      = \frac{\pi}{2}\; \left. _2F_1\left(
      \begin{array}{c} \frac{1}{2}, \; \frac{1}{2} \\ 1 \end{array}
      \right| k^2 \right) \; ,
\\
\label{elliptic_Pi}
\Pi(c,k) &=& \int\limits_0^1 
\frac{\mbox{d}t}{(1-ct^2)\sqrt{(1-t^2)(1-k^2t^2)}}
      = \frac{\pi}{2}\; 
      F_1\left({\textstyle{\frac{1}{2}}}; \; 1, \; 
      {\textstyle{\frac{1}{2}}}; \; 1 |\; c,\; k^2 \right),
\end{eqnarray}
where $F_1$ is the Appell hypergeometric function of two arguments.

The Jacobian zeta function, $Z(\beta,k)$, is defined through
\begin{equation}
\label{def_Zeta}
K(k)\; Z(\beta,k) = K(k)\; E(\beta,k) - E(k)\; F(\beta,k) .
\end{equation}
We will assume that $0\leq k < 1$. (In the limit $k\to1$, 
$F(\beta,k)$ and $K(k)$ are singular.) From the definition~(\ref{def_Zeta}) 
it is obvious that $Z\left(\frac{\pi}{2},k\right)=0$.
Moreover, using symmetry properties of $E(\beta,k)$ and $F(\beta,k)$,
we get
\begin{equation}
\label{Zeta_continued}
Z\left({\textstyle{\frac{\pi}{2}}}+\delta, k\right) =
- Z\left({\textstyle{\frac{\pi}{2}}}-\delta, k\right) \; . 
\end{equation}

To represent the elliptic functions $\Pi(\alpha_i,k)$ occurring in 
Eq.~(\ref{bbbb}) in terms of $Z$ functions, we can use
\begin{equation}
\label{caseIII}
\Pi(\alpha_i^2,k) = K(k) 
+\frac{\alpha_i \; K(k) \; 
Z(\beta_i,k)}{\sqrt{(1-\alpha_i^2)(k^2-\alpha_i^2)}},
\end{equation}
with 
$\beta_i = \arcsin(\alpha_i/k)$.
Eq.~(\ref{caseIII}) corresponds to case III on p.~229 of \cite{Byrd},
when $0\le \alpha_i^2<k^2$.

The following addition formulae from Ref.~\cite{Byrd} 
(p.~34, Eq.~(142.01)) are needed: 
\begin{equation}
\label{Z_addition}
Z(\beta_1,k) \pm Z(\beta_2,k) = Z(\varphi_{\pm},k)
\pm k^2 \sin\beta_1 \sin\beta_2 \sin\varphi_{\pm} ,
\end{equation}
where the angles
\begin{equation}
\varphi_{\pm} = 2 \arctan\left[
\frac{\sin\beta_1 \sqrt{1-k^2 \sin^2\beta_2}
      \pm \sin\beta_2 \sqrt{1-k^2 \sin^2\beta_1}}
    {\cos\beta_1 + \cos\beta_2} \right] \; .
\end{equation}
In fact, the angles $\varphi_{-}$ and $\varphi_{+}$ correspond
to the angles $\varphi_1$ and $\varphi_2$ 
(see Eqs.~(\ref{cos_phi})--(\ref{sin_phi})), respectively.
Moreover, using the same addition formula~(\ref{Z_addition})
with $\beta_{1,2}$ substituted by $\varphi_{1,2}$, we get
the symmetric connection~(\ref{Z+Z+Z}) between
$Z(\varphi_i,k)$ ($i=1,2,3$).

To derive the result given in Eq.~(\ref{sqrt_Z3}), one can use
the integral tables of Ref.~\cite{Byrd}, along with the
following relation:
\begin{equation}
\label{141.01}
2Z(\beta_1,k) = -Z(\varphi_3,k)
+\frac{2k^2\sin^3\beta_1\cos\beta_1\sqrt{1-k^2\sin^2\beta_1}}{1-k^2\sin^4\beta_1}
\; .
\end{equation}
It corresponds to the last two lines of Eq.~(141.01) on p.~33 of Ref.~\cite{Byrd},
where $\varphi\leftrightarrow\pi-\varphi_3$.


\end{document}